\documentclass{article}

\usepackage{amsmath}
\usepackage{PRIMEarxiv}
\usepackage{natbib} 
\usepackage[utf8]{inputenc} 
\usepackage[T1]{fontenc}    
\usepackage{hyperref}       
\usepackage{url}            
\usepackage{booktabs}       
\usepackage{amsfonts}       
\usepackage{nicefrac}       
\usepackage{microtype}      
\usepackage{lipsum}
\usepackage{fancyhdr}       
\usepackage{graphicx}       
\graphicspath{{media/}}     
\usepackage{hyperref}
\hypersetup{
 pdfauthor={Md Saiful Islam, Rahul Bhadani},   
 pdftitle={Resilient Composite Control for Stability Enhancement in EV Integrated DC Microgrids},   
 pdfsubject={Power Systems, Control Systems, Power Electronics}, 
 pdfkeywords={DC microgrid, constant power load, virtual
capacitor, composite controller},
 pdfcreator={Md Saiful Islam}, 
 pdfproducer={Md Saiful Islam}, 
 colorlinks=true,
 filecolor=green, 
 citecolor = blue,       
 urlcolor=red, 
 }
\usepackage{mathpazo}
\usepackage{microtype}
\usepackage{caption}
\title{Resilient Controller Design with
Exponential Reaching Law for Enhanced Load Frequency Stability in Multi-Area Interconnected Microgrids 
}
\usepackage{times}
\usepackage{threeparttable}
\author{
  Md Saiful Islam, Rahul Bhadani \\
  AI, Autonomy, Resilience, Control (AARC) Lab\\
  Electrical \& Computer Engineering\\
  The University of Alabama in Huntsville\\
  301 Sparkman Drive, Huntsville, AL 35899, USA\\
  \texttt{mi1499@uah.edu, rahul.bhadani@uah.edu} \\
}

\begin{document}
\maketitle

\begin{abstract}
We present a load frequency control strategy deploying a decentralized robust global integral terminal sliding mode control (GITSMC) method to maintain stable frequency and tie-line power in multi-area interconnected microgrids with aggregated uncertainties. To achieve this, firstly, we have developed a mathematical model of the multi-area interconnected system incorporating disturbances from solar photovoltaic (PV), wind turbine (WT) generation and load demand, as aggregated uncertainties. Secondly, we have designed a global integral terminal sliding surface with an exponential reaching law for each area to enhance system dynamic performance and suppress chattering within a finite time. Thirdly, the overall stability of the closed-loop system is analyzed using the Lyapunov stability theorem. Finally, extensive simulations are conducted on the IEEE 10-generator New England 39-bus power system, including load disturbances and variable PV and WT generation. The results demonstrate the performance of the proposed GITSMC approach, achieving approximately 94.9\% improvement in ITSE and 94.4\% improvement in ISE, confirming its superior accuracy and dynamic performance compared to the existing controller.

\end{abstract}


\section{Introduction}
\begin{figure}[h]
	\centering
	\includegraphics[width=0.9\textwidth]{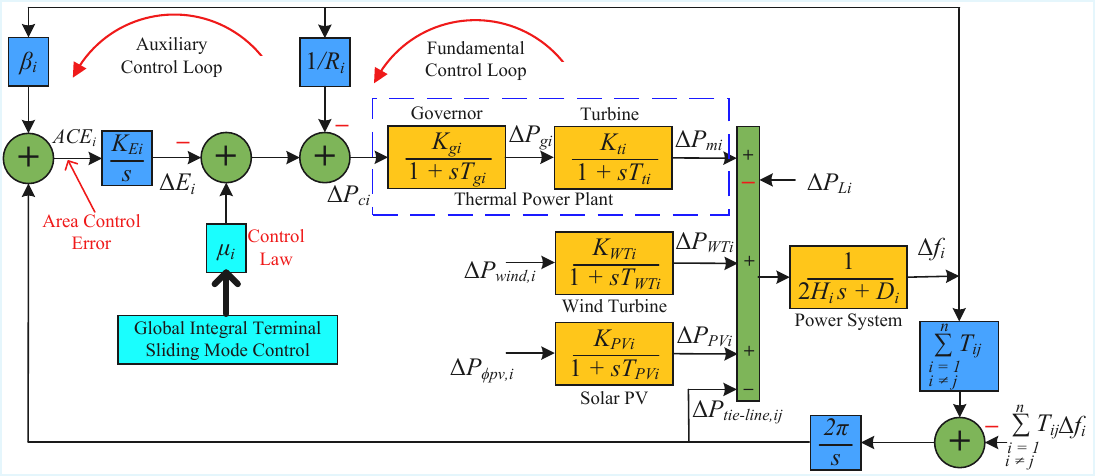}
	\caption{Multi-area interconnected microgrids with thermal power plant along with PV and wind generators.}
	\label{Fig1}
\end{figure}

Rising concerns over climate change and fossil fuel depletion have accelerated the global shift toward renewable-based power systems. Wind turbine (WT) and photovoltaic (PV) power generation, being clean and sustainable, play a major role in electric power production~\citep{islam2024enhanced, islam2025enhanced, islam2025marine}. However, their intermittent and weather-dependent nature introduces notable voltage and frequency variations~\citep{feng2025online}. In large interconnected grids, high renewable penetration intensifies operational uncertainty, reduces system inertia, and heightens vulnerability to generation–load imbalances~\citep{ghatuari2025coordinated}. These fluctuations can cause significant frequency deviations, leading to grid destabilization. Although primary control actions, such as turbine governors, respond to disturbances, they often fail to eliminate steady-state errors or suppress oscillations. Therefore, the load frequency control (LFC) is required to maintain stable tie-line (transmission line connecting two or more power systems~\citep{machowski1997power}) power and frequency. Recent research emphasizes robust LFC strategies to enhance transient stability~\citep{shangguan2025dissipativity, biswas2025dynamic} in renewable-integrated multi-area interconnected microgrids (MAIMGs) as presented in Figure~\ref{Fig1}. When several geographically dispersed AC and DC microgrid systems are linked together through interconnections, they collectively form what is known as a \textbf{multi-area interconnected microgrid}~\citep{John_2017}.

The sliding mode controller (SMC) has attracted attention for its robustness in LFC under parameter variations and external disturbances. A robust $H_{\infty}$ based SMC was proposed in~\citep{sun2017robust} for MAIMGs, addressing time-delay disturbances. However, unmodeled dynamics and external disturbances were not considered. To overcome this, an extended disturbance observer-based SMC was introduced in~\citep{guo2025observer}, treating tie-line and load variations as disturbances. Yet, parameter variations and renewable generation impacts remained unaddressed. To address these,~\citep{guha2023improved} proposed an observer-based fractional-order SMC for hybrid wind–diesel systems, though the use of a discontinuous signum function caused oscillations, limiting practical feasibility.

Fractional-order terminal SMC achieves uniform voltage regulation under wide load and input variations~\citep{yu2025predictor}, while global fast terminal SMC improves convergence by adding linear terms, enhancing dynamic response~\citep{qian2025load}. Adaptive TSMC integrated with energy storage was proposed in~\citep{zhang2025coordinated}, but its effectiveness under mismatched disturbances from intermittent renewable sources was not evaluated. Similarly,~\citep{ngamroo2016design} did not fully analyze the effects of disturbances induced by renewable sources. Moreover, the widespread deployment of energy storage faces design complexities and higher costs. To address these limitations,~\citep{alhassan2025influence} suggested a robust, chattering-free adaptive higher-order integral SMC for hydro-based LFC, but neglecting other renewable sources like solar PV and WT is impractical. Higher-order integral SMCs also involve complex stability proofs and higher-order derivatives, increasing noise. Integral terminal SMC improves convergence, reduces chattering, and enhances dynamics~\citep{islam2024enhanced}.

Building upon the benchmark multi-area LFC models established in~\citep{sharifi2008load, sahu2015design, roy2024robust, roy2024multi}, this paper proposes a robust decentralized global integral terminal sliding mode controller (GITSMC) to achieve improved tie-line power and frequency regulation in multi-area interconnected systems with integrated WT and PV power generation.

The main contributions are as follows:
\begin{enumerate}
    \item Development of a multi-area interconnected LFC model incorporating solar and wind sources, considering parametric uncertainty and external disturbances as lumped uncertainties.
    \item Design of a decentralized robust GITSMC in conjunction with an exponential reaching law that suppresses chattering, avoids singularity while minimizing steady-state error, and ensures fast finite-time convergence.
    \item Comparative analysis with the classical proportional-integral (PI) controller, demonstrating the superior performance of the proposed controller.
    \item Validation on the IEEE 10 generators 39-bus New England power system under varying generation and load conditions.
\end{enumerate}

\section{Modeling of Multi-Area Interconnected Microgrids}

Figure~\ref{Fig1} illustrates the $i^{th}$ area of an interconnected power system comprising a PV power generator, a thermal power plant, a wind power generator, and loads. The tie-line power deviation ($\Delta P_{\text{tie-ij}}$) and the frequency deviation ($\Delta f_i$) are adjusted by controlling the mechanical output power ($\Delta P_{mi}$) in response to load ($\Delta P_{Li}$), solar PV ($\Delta P_{\text{PVi}}$), wind ($\Delta P_{\text{WTi}}$), and tie-line power variations, utilizing the speed governor in the primary control loop. Since the primary controller cannot fully restore frequency, a decentralized robust GITSMC is proposed to mitigate tie-line power as well as frequency deviations. For simplicity, the $i^{th}$ area is modeled as~\citep{ sahu2015design, roy2024robust, roy2024multi, yang2021disturbance}:
\begin{eqnarray} 
	\begin{aligned}\label{eq:CM1}
			&\Delta\dot{{P}}_{\text{tie-ij}} ={2\pi}\left(\sum_{i=1, i\neq j}^{n}T_{ij}\left( \Delta{f}_{i}\left( t\right)-\Delta{f}_{j}\left( t\right)\right)\right) \\
		&\Delta{\dot{f}_i}={f}_{i}\left( t\right)+{P}_{mi}\left( t\right)+{P}_{PVi}+{P}_{WTi}\left(t\right)+{P}_{tie-ij}\left(t\right)+{P}_{Li}\left( t\right)\\	
	&\Delta{\dot{P}_{mi}}=-\frac{K_{ti}}{T_{ti}}\left(\Delta{P}_{mi}\left( t\right)-\Delta{P}_{gi}\left(t\right)\right)\\
	&\Delta{\dot{E}_{i}}=K_{Ei}\left(\beta_{i}\Delta{f}_{i}\left(t\right)+\Delta{P}_{\text{tie-ij}}\left( t\right)\right)\\
	&\Delta{\dot{P}_{gi}}=\frac{K_{gi}}{T_{gi}}\left(\mu_{i}-\Delta E_{i}-\frac{\Delta f_i}{R_i}-\Delta P_{gi}\right) \\
	&\Delta{\dot{P}_{\text{PVi}}}=\frac{1}{T_{\text{PVi}}}\left(-\Delta{P}_{\text{PVi}}+\Delta{P}_{\phi i} K_{\text{PVi}}\right)\\
	&\Delta{\dot{P}_\text{{WTi}}}=\frac{1}{T_{\text{WTi}}}\left( K_{wi}\Delta{P}_\text{{wind-i}}-\Delta{P}_\text{{WTi}}\right)\\
	\end{aligned} 
\end{eqnarray}
in which ${f}_{i}\left( t\right)=-\frac{D_i}{2H_i}\Delta{f}_{i}\left( t\right)$, ${P}_{mi}=\frac{1}{2H_i}\Delta{P}_{mi}\left( t\right)$, ${P}_{PVi} = \frac{1}{2H_i}\Delta{P}_{PVi}$, ${P}_{WTi} = \frac{1}{2H_i}\Delta{P}_{WTi}\left(t\right)$, ${P}_{tie-ij} = - \frac{1}{2H_i}\Delta{P}_{tie-ij}\left(t\right)$, and ${P}_{Li} = -\frac{1}{2H_i}\Delta{P}_{Li}\left( t\right)$. $\beta_i$, $R_i$, $T_{ij}$, $D_i$, $H_i$, $K_{ti}$, $T_{ti}$, $K_{gi}$, $T_{gi}$, $T_{PVi}$, and $T_{wi}$ are system coefficients and time constants. $\Delta f_i$, $\Delta f_j$, $\Delta P_{mi}$, $\Delta P_{PVi}$, $\Delta P_{WTi}$, $\Delta P_{Li}$, $\Delta P_{gi}$, and $\Delta P_{\text{wind-}i}$ are deviations or variations, $\mu_i$ is control signal, details are presented in~\citep{ sahu2015design, roy2024robust, roy2024multi, yang2021disturbance}.
Now, Equation~\eqref{eq:CM1} can be rewritten in state-space form as:
\begin{eqnarray} 
	\begin{aligned}\label{eq:CM2}
		\dot{x}_{i}=A_{i}x_{i}\left(t\right)+B_{i}u_{i}\left(t\right)+\sum_{i=1, i\neq j}^{n}A_{ij}x_{j}\left(t\right)+\psi_{i}\Delta{P}_{di}
	\end{aligned}
\end{eqnarray} 
in which $x_{i}=\left[\Delta{P}_\text{{tie-ij}}~\Delta{f}_{i}~\Delta{P}_{mi}~\Delta{E}_{i}~\Delta{P}_{gi}~\Delta{P}_\text{{PVi}}~\Delta{P}_\text{{WTi}} \right]^{T}$ is the state variables, $
A_{ij} =
\begin{pmatrix}
a_{11} & 0_{1 \times 6} \\
0_{6 \times 1} & 0_{6 \times 6}
\end{pmatrix}
$, with $a_{11}=-2\pi\sum_{i=1, i\neq j}^{n}T_{ij}$,
\begin{equation*}\nonumber
	A_{i} = 
		\setlength{\arraycolsep}{0.5pt}
	\renewcommand{\arraystretch}{1.3}
	\begin{pmatrix}
		0 & \gamma_i & 0 & 0 & 0 & 0 & 0 \\
		-\frac{1}{2H_{i}} &-\frac{D_i}{2H_{i}}& \frac{1}{2H_{i}} & 0 &0 &\frac{1}{2H_{i}}&\frac{1}{2H_{i}} \\
		0& 0& -\frac{K_{ti}}{T_{ti}}& 0 & \frac{K_{ti}}{T_{ti}}& 0 & 0  \\
		K_{Ei}& K_{Ei}\beta_{i} & 0 & 0 & 0 & 0 & 0\\
		0 & -\frac{K_{gi}}{R_{i}T_{gi}} & 0 & -\frac{K_{gi}}{T_{gi}}& -\frac{K_{gi}}{T_{gi}} & 0 & 0\\
		0 & 0 & 0 & 0 & 0 & -\frac{1}{T_\text{{PVi}}} & 0\\
		0 & 0 & 0 & 0 & 0 & 0 & -\frac{1}{T_\text{{WTi}}}
	\end{pmatrix}
\end{equation*}
with $\gamma_i=2\pi\sum_{i=1, i\neq j}^{n}T_{ij}$,
$B_{i}=\left[0~~0~~0~~0~~\frac{K_{gi}}{T_{gi}}~~0~~0\right]^{T}$, \begin{equation}\nonumber
	\psi_{i} = 
	\begin{pmatrix}
		0 & 0 & 0 \\
	-\frac{1}{2H_{i}} & 0 & 0\\
	0 & 0 & 0 \\
	0 & 0 & 0 \\
	0 & 0 & 0 \\
	0 & \frac{K_{\text{PVi}}}{T_{\text{PVi}}} & 0 \\
		0 & 0 & \frac{K_{\text{WTi}}}{T_{\text{WTi}}} 
	\end{pmatrix}
\end{equation}
and $\Delta{P}_{di}=\left[\Delta{P}_{Li}~~\Delta{P}_{\phi i}~~\Delta{P}_{\text{wind-i}}\right]^{T}$.

Matrices $A_i$, $A_{ij}$, $B_i$, and $\delta_i$ in Equation~\eqref{eq:CM2} are assumed known, but real-world variations in frequency, tie-line power, and generation, along with load fluctuations and system reconfigurations, introduce modeling uncertainties. Thus, the power system dynamics under uncertainty is~\citep{roy2024robust, roy2024multi}:
\begin{eqnarray}
	\begin{aligned}\label{eq:CM3}
		\dot{x}_{i}&=\left(A_{0i}+\Delta{A}_{i}\right) x_{i}+\left( B_{0i}+\Delta{B}_{i}\right)\mu_{i}+\sum_{i=1, i\neq j}^{n}\left( A_{0ij}+\Delta{A}_{ij}\right) x_{j}+\left( \psi_{0i}+\Delta{\psi}_{i}\right) \Delta{P}_{di}
	\end{aligned}
\end{eqnarray}
where $A_{0i}, B_{0i}, A_{0ij}$, and $\psi_{0i}$ stand the nominal system matrices, and $\Delta A_i, \Delta B_i, \Delta A_{ij}, \Delta \psi_i$ denote the uncertainties. Then, Equation~\eqref{eq:CM3} becomes:
\begin{eqnarray}
	\begin{aligned}\label{eq:CM4}
		\dot{x}_{i}=A_{i}x_{i}\left( t\right) +B_{i}\mu_{i}\left( t\right) +\sigma_{i}
	\end{aligned}
\end{eqnarray}
where $\sigma_{i}=\Delta{A}_{i}+\Delta{B}_{i}\Delta{\mu}_{i}+\sum_{i=1, i\neq j}^{n}\left( A_{ij}+\Delta{A}_{ij}\right) x_{j}+\left( \psi_{i}+\Delta{\psi}_{i}\right) \Delta{P}_{di}$ represents the lumped perturbation. In Equation~\eqref{eq:CM4}, the dimensions are $x_{i}\in R^{n \times 1}, A_{i}\in R^{n \times n}, B_{i}\in R^{n \times m}, \mu_{i}\in R^{m \times 1},$ and $\sigma_{i} \in R^{n \times 1}$.
\\
\textit{\textbf{Assumption 1.} The lumped disturbance is bounded, i.e., \(\|\sigma_i\| \leq \zeta_i\), where \(\zeta_i > 0\) is known.}
\section{Design of the Proposed Robust GITSMC with an Exponential Reaching Law}
In this section, a robust GITSMC is developed to guarantee finite-time convergence of the system output to equilibrium. To mitigate lumped perturbations, enhance transient performance, eliminate steady-state error, and reduce chattering, a global integral terminal sliding surface~\citep{islam2025stability} ($\Theta_i(t) \in \mathbb{R}^{m \times 1}$) is proposed as follows: 
\begin{eqnarray}
	\begin{aligned}\label{eq:CD1}
\Theta_{i}\left(t\right)=\vartheta_{i}x_{i}\left(t\right) +\vartheta_{i}\lambda_{1i}\int x\left( t\right) dt+\vartheta_{i}\lambda_{2i}\int\left(x^{\alpha_{i}}\left( t\right) \right) dt
	\end{aligned}
\end{eqnarray}
where $\lambda_{i1}$, $\lambda_{i2}$ and $\alpha_i$ ($1 < \alpha_i < 2$) are user-defined constant. The matrix $\vartheta_i \in \mathbb{R}^{m \times n}$ should be selected to ensure that the product $\vartheta_i B_{0i}$ is non-singular. The time derivative of $\Theta_i(t)$ incorporating Equation~\eqref{eq:CM4} is:
\begin{eqnarray}
	\begin{aligned}\label{eq:CD2}
        		\dot{\Theta}_{i}\left(t\right)&=\vartheta_{i}A_{0i}x_{i}\left( t\right) +\vartheta_{i}B_{0i}\mu_{i}\left( t\right) +\vartheta_{i}\sigma_{i} +\vartheta_{i}\lambda_{1i}x\left( t\right)+\vartheta_{i}\lambda_{2i}\left(x^{\alpha_{i}}\left( t\right) \right) 
	\end{aligned}
\end{eqnarray}
Setting $\dot{\Theta}_{i}\left(t\right)=0$, we have: 
\begin{eqnarray}
	\begin{aligned}\label{eq:CD3}
        0=\vartheta_{i}A_{0i}x_{i}\left( t\right) +\vartheta_{i}B_{0i}\mu_{i}\left( t\right) +\vartheta_{i}\sigma_{i} +\vartheta_{i}\lambda_{1i} x\left( t\right) +\vartheta_{i}\lambda_{2i}\left(x^{\alpha_{i}}\left( t\right) \right)
	\end{aligned}
\end{eqnarray}
Thus, neglecting the uncertainties in Equation~\eqref{eq:CD3}, the equivalent control law is: 
\begin{eqnarray}
	\begin{aligned}\label{eq:CD4}
    	\mu_{eqi}\left( t\right) =-\frac{1}{\vartheta_iB_{0i}}\left(\vartheta_iA_{0i}x_{i}\left( t\right)+\vartheta_{i}\sigma_{i}+ \vartheta_{i}\lambda_{1i} x\left( t\right)  +\vartheta_{i}\lambda_{2i}x^{\alpha_{i}}\left( t\right)\right) 
	\end{aligned}
\end{eqnarray}
An exponential reaching law~\citep{islam2025stability} is deployed to suppress the chattering and achieve fast finite-time convergence, as follows: 
\begin{eqnarray}
	\begin{aligned}\label{eq:CD5}
		\mu_{swi}\left( t\right) =-\frac{1}{\vartheta_iB_{0i}}\left( \eta_{1i}\Theta_{i}\left( t\right) +\eta_{2i}\frac{\Theta_{i}\left( t\right) }{||\Theta_{i}||}\right) 
	\end{aligned}
\end{eqnarray}
where $\eta_{1i}>0, \eta_{2i}>0 \in \mathbb{R}^{1 \times m}$ are user-defined switching matrices. Accordingly, the overall control law $\mu_i(t)$ based on the SMC principle is given by: \vspace{-11pt}
\begin{eqnarray}
	\begin{aligned}\label{eq:CD6}
		\mu_{i}\left( t\right) =\mu_{eqi}\left( t\right)+\mu_{swi}\left( t\right)
	\end{aligned}
\end{eqnarray}
The overall control law incorporating Equation~\eqref{eq:CD4} and Equation~\eqref{eq:CD5}  is: 
\begin{eqnarray}
\begin{aligned}\label{eq:CD7}
        		\mu_{i}\left(t\right)=&-\frac{1}{\vartheta_iB_{0i}}\left(\vartheta_iA_{0i}x_{i}+\vartheta_{i}\lambda_{1i} x\left( t\right) +\vartheta_{i}\sigma_{i}+\vartheta_{2i}\lambda_{i}x^{\alpha}_{i}\right)-\frac{1}{\vartheta_iB_{0i}}\left(  \eta_{1i}\Theta_{i}+\eta_{2i}\frac{\Theta_{i}}{||\Theta_{i}||}\right)
\end{aligned}
\end{eqnarray}
Now, Equation~\eqref{eq:CD2} becomes:  \vspace{-11pt}  
\begin{eqnarray}
	\begin{aligned}\label{eq:CD8}
		\dot{\Theta}_{i}\left(t\right)=-\eta_{1i}\Theta_{i}\left( t\right) -\eta_{2i}\frac{\Theta_{i}\left( t\right) }{||\Theta_{i}||}+\vartheta_{i}\sigma_{i}\left( t\right) 
	\end{aligned}
\end{eqnarray}
The Lyapunov control function is chosen as follows to verify the overall system stability: 

\begin{eqnarray}
	\begin{aligned}\label{eq:CD9}
L_{i}\left( t\right) =\frac{1}{2}\Theta^{2}_{i}
	\end{aligned}
\end{eqnarray}
\textit{\textbf{Remark 1.}  If the Lyapunov function $L_i(t)$ is positive‐definite, then the system is said to be asymptotically stable and should meet the conditions: $L_i(t) > 0 \ \forall\ \Theta_i(t) \neq 0, \quad \dot{L}_i(t) \le 0,\ \text{and} \quad L_i(0) = 0$}.
\\
The time derivative of $L_i(t)$ based on Equation~\eqref{eq:CD8} is: 
\begin{eqnarray}
	\begin{aligned}\label{eq:CD10}
		\dot{L}_{i}\left( t\right) =\Theta_{i}\left( t\right)\left( -\eta_{1i}\Theta_{i}\left( t\right) -\eta_{2i}\frac{\Theta_{i}\left( t\right) }{||\Theta_{i}||}+\vartheta_{i}\sigma_{i}\left( t\right) \right) 
	\end{aligned}
\end{eqnarray}
which becomes: 
\begin{eqnarray}
	\begin{aligned}\label{eq:CD11}
		\dot{L}_{i}\left( t\right)\leq -||\eta_{1i}\Theta^2_{i}||-\frac{||\eta_{2i}\Theta^2_{i}||}{||\Theta_{i}||}+||\vartheta_{i}\sigma_{i}\Theta_{i}||
	\end{aligned}
\end{eqnarray}
We can rewrite Equation~\eqref{eq:CD11} applying the inequality $||mn|| \leq ||m||||n||$ as:
\begin{eqnarray}
	\begin{aligned}\label{eq:CD12}
		\dot{L}_{i}\left( t\right)&\leq -||\eta_{1i}||||\Theta^2_{i}||-\frac{||\eta_{2i}||||\Theta^2_{i}||}{||\Theta_{i}||}+||\vartheta_{i}||||\sigma_{i}||||\Theta_{i}||
	\end{aligned}
\end{eqnarray}
which becomes: 
\begin{eqnarray}
	\begin{aligned}\label{eq:CD13}
		\dot{L}_{i}\left( t\right)&\leq -||\eta_{1i}||||\Theta^2_{i}||-||\eta_{2i}||||\Theta_{i}||+||\vartheta_{i}||||\sigma_{i}||||\Theta_{i}||
	\end{aligned}
\end{eqnarray}
Under \textit{Assumption 1}, Equation~\eqref{eq:CD13} simplifies to: 
\begin{eqnarray}
	\begin{aligned}\label{eq:CD14}
		\dot{L}_{i}\left( t\right)&\leq -||\eta_{1i}||||\Theta^2_{i}||-||\Theta_{i}||\left( ||\eta_{i}||-||\vartheta_{i}||\mu_{i}\right) 
	\end{aligned}
\end{eqnarray} 
As $||\eta_{1i}|| > 0$ and $||\eta_{2i}|| \geq ||\vartheta_i||\zeta_i$, we have $\dot{L}_i(t) \leq 0$. Therefore, from Equation~\eqref{eq:CD14}, the overall LFC system under the control law Equation~\eqref{eq:CD7} is stable, and the sliding surface converges to zero in finite time despite aggregated uncertainties. The finite time convergence analysis is briefly given in Appendix~\ref{app:A}.


The next section presents numerical and comparative results evaluating the proposed control method.
\begin{figure} [h]
	\vspace{-0.1cm}
	\centering 
	\includegraphics[width=0.9\textwidth]{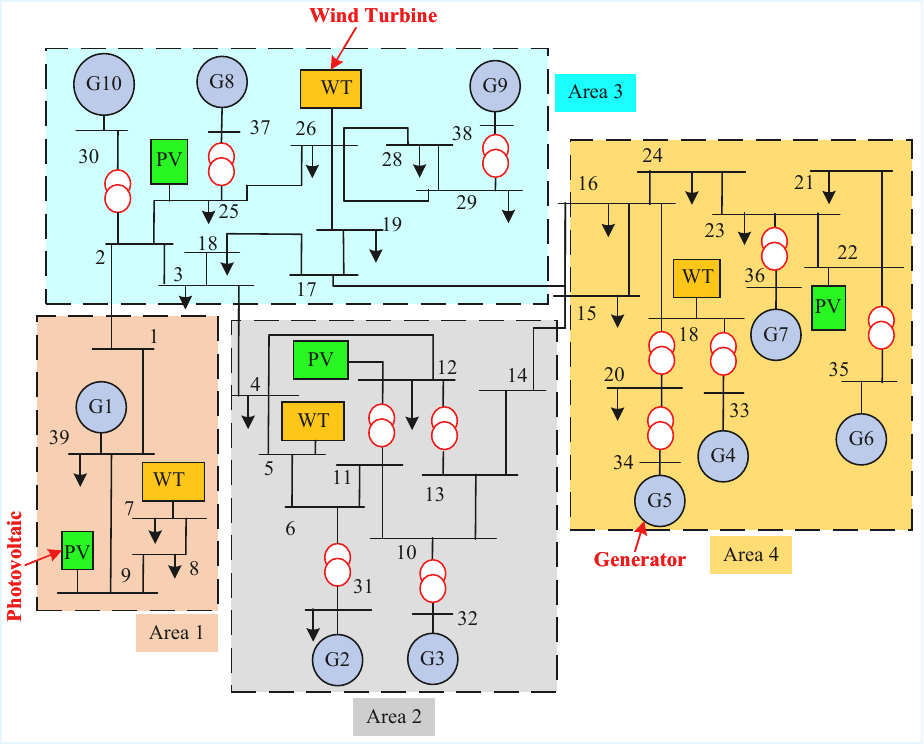}
	\caption{Single line illustration of the IEEE 10-generator New England 39-bus power system~\citep{roy2024robust, roy2024multi, IEEE39bus, yan2020multi}.}
    \label{IEEE39}
\end{figure}


 \renewcommand{\arraystretch}{1.35}
\begin{table} [h]
\centering
{\footnotesize
	\caption{Parameters of the IEEE 39-bus (New England) power network~\citep{roy2024robust, roy2024multi, IEEE39bus, yan2020multi}}
	\label{tab1}
	\begin{tabular}{ |c|c|c|c|c|c|c|c|c|c|c|c} 
		\hline
		Area & Generator&S  & $T_t$ & $T_g$  & R & H & $T_{ij}$\\
		\hline
		1 & $G_1$ & 1000 & 0.3742 & 0.0804 & 0.0471  & 10 & $T_{13}$=1.3272\\
		\hline
		& $G_2$ & 520.81 & 0.3888 & 0.0774 & 0.0541  & 6.06 & $T_{23}$=0.2959\\

		2 & $G_3$ & 650 & 0.3645 & 0.0748 & 0.0518  & 7.16 & $T_{24}$=0.6128\\
		\hline	
		& $G_4$ & 632 & 0.3707 & 0.0759 & 0.0540  & 5.72 & $T_{31}$=1.3272\\

		3& $G_5$ & 508 & 0.3770 & 0.0729 & 0.0470  & 5.20 & $T_{32}$=0.2959\\

		& $G_6$ & 650 & 0.4316 & 0.0791 & 0.0459  & 6.96 & $T_{34}$=0.3959\\

		& $G_7$ & 560 & 0.3657 & 0.0722 & 0.0481  & 5.28 & \\
		\hline
		& $G_8$ & 540 & 0.3665 & 0.0805 & 0.0484  & 4.86 & $T_{42}$=0.6128\\
		
		4 & $G_9$ & 830 & 0.4222 & 0.0737 & 0.0479  & 6.90 & $T_{43}$=0.3959\\
		
		& $G_{10}$ & 250 & 0.4324 & 0.0852 & 0.0525  & 8.40 & \\
		\hline
	\end{tabular}
}
		\begin{tablenotes} \centering
		\item {\scriptsize $D=$1 for all the generators}
	\end{tablenotes}
\end{table}
\section{Results and Analysis}

The IEEE 10-generator New England 39-bus power network~\citep{roy2024robust, roy2024multi, IEEE39bus, yan2020multi}, as given in Figure~\ref{IEEE39}, is deployed to evaluate the efficacy of the constructed decentralized GITSMC scheme. This widely adopted test system is frequently employed in dynamic stability analyses and is partitioned into four interconnected subareas. This IEEE 39-bus system is subdivided into four different areas. Area 1 comprises generator $G_1$. Area 2 includes generators $G_2$ and $G_3$, while area 3 consists of generators $G_8$, $G_9$, and $G_{10}$. The generators $G_4$–$G_7$ are assigned to area 4. Each area is also equipped with a PV unit, a WT unit, and step load disturbances. The power system’s net installed capacity amounts to 6408 MVA, with load demands of 959 MW, 1867 MW, 2350 MW, and 1707.5 MW corresponding to areas 1–4, respectively. As shown in Figure~\ref{IEEE39}, the tie-line links between the subareas are arranged as follows: area 1 is connected to area 3 through buses 1–2; area 2 is linked to area 3 via buses 3–4; area 2 connects to area 4 through buses 14–15; and area 3 is tied to area 4 at buses 16–17. Four WT units, each rated at 100 MW and totaling 400 MW, are connected to buses 5, 7, 18, and 26. Similarly, four solar PV units, also rated at 100 MW each and totaling 400 MW, are integrated at buses 9, 12, 22, and 25. Consequently, the combined renewable generation from WT and PV sources equals 800 MW, corresponding to a per-unit value of 0.8. Moreover, Table~\ref{tab1} presents the tie-line synchronizing torque coefficients for this four-area power system~\citep{jin2023delay}. Table~\ref{tab2} provides an overview of the parameters utilized in developing the proposed four-area LFC model. Using the data from Table~\ref{tab2}, the matrices for the interconnected power systems in the four areas are calculated, and they are presented in Appendix~\ref{app:B}. The simulation data for a PV unit is $T_{\text{PV}}=1.8$, and $K_{\text{PV}}=1$; and for a WT is $T_{\text{WT}}=1.5$, and $K_{\text{WT}}=1$.
The effectiveness of the proposed controller is assessed by incorporating step load disturbances along with the presence of WT and PV power generators in all areas. In this paper, the performance of the designed decentralized robust GITSMC is compared with that of a classical PI controller. For the simulation the following control parameters of the proposed GITSMC are used:~$\alpha_1=\alpha_2=\alpha_3=\alpha_4= 1.7$, $\lambda_1=\lambda_2=\lambda_3=\lambda_4=24$, $\eta_{11}=\eta_{12}=\eta_{13}=\eta_{14}
=\left[1.498~~1.991~~  0.637~~  0.728~~  0.1582~~  0.248~~  0.239\right]$,
and $\eta_{21}=\eta_{22}=\eta_{23}=\eta_{24}
=\left[2.498~~  2.991~~  0.737~~  0.728~~  0.258~~  0.148~~  0.339\right]$. 

 \renewcommand{\arraystretch}{1.5}
\begin{table}[!t]
\centering
{\footnotesize
\caption{Model parameters for the load frequency control scheme~\citep{roy2024robust, roy2024multi, IEEE39bus, yan2020multi}}
\label{tab2}
\begin{tabular}{|c|c|c|c|c|c|c|}
\hline
Area & $T_{\text{t-eqv}}$ & $T_{\text{g-eqv}}$ & $R_{\text{eqv}}$ & $\beta$ & $H_{\text{eqv}}$ & $T_{ij}$ \\
\hline
1 & 0.3742 & 0.0804 & 0.0471 & 22.2314 & 10 & $T_{13}$\\
\hline
2 & 0.3766 & 0.0760 & 0.0528 & 6.6706 & 19.9394 & $T_{23}, T_{24}$\\
\hline
3 & 0.3862 & 0.0750 & 0.0486 & 21.5761 & 6.6706 & $T_{31}, T_{32}, T_{34}$\\
\hline
4 & 0.4070 & 0.0798 & 0.0487 & 21.5339 & 6.4515 & $T_{42}, T_{43}$\\
\hline
\end{tabular}
}
\begin{tablenotes}\centering
\item {\scriptsize $T_{13}=T_{31}=1.3272$, $T_{23}=T_{32}=0.2959$, $T_{24}=T_{42}=0.6128$, $T_{43}=0.3959$, and $D_{\text{eqv}}=1$ in all areas.}
\end{tablenotes}
\end{table}

\begin{figure}[h]
	\hspace{-0.5cm}
	\vspace{-0.1cm}
	\centering
	\includegraphics[width=1\textwidth]{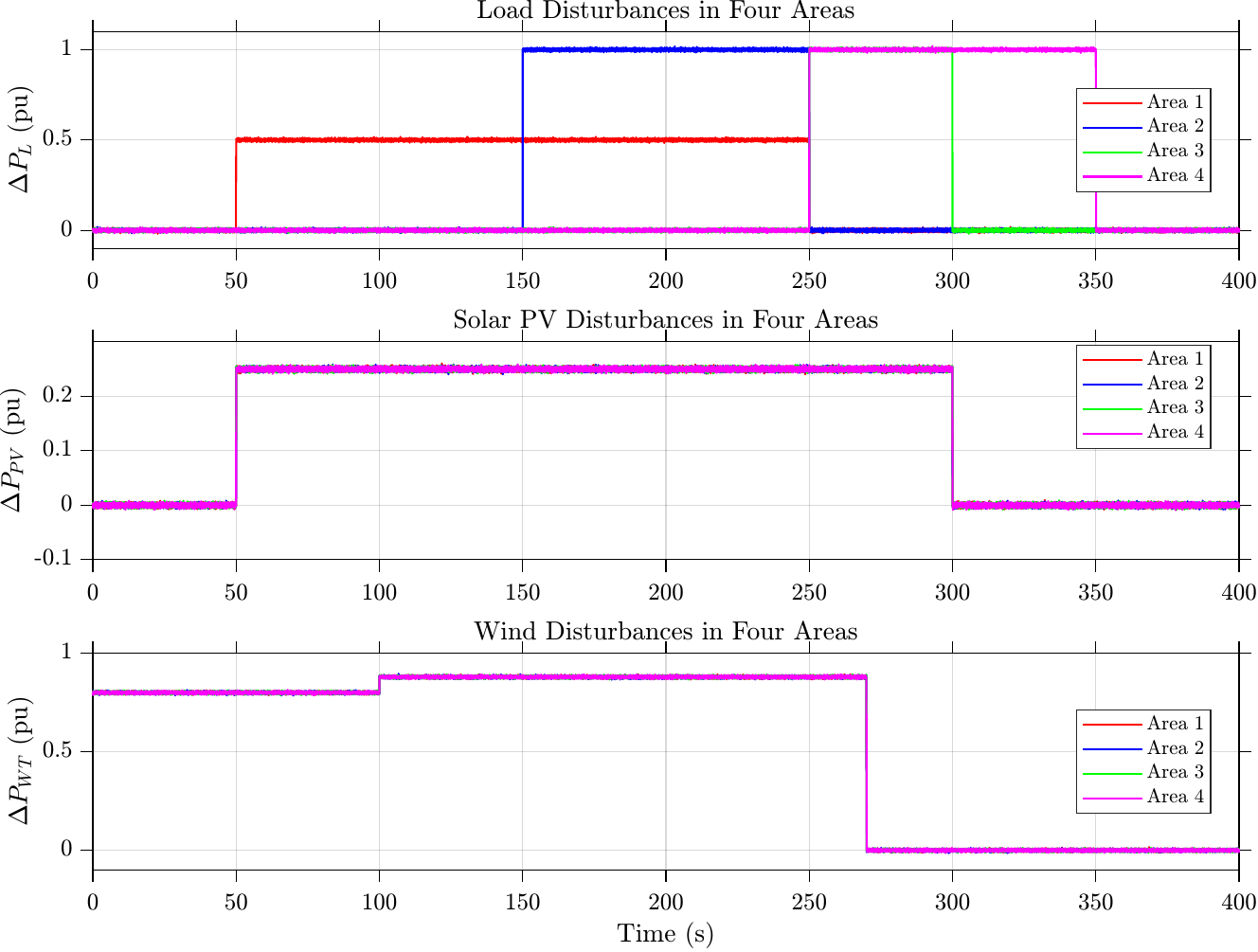}
	\caption{Step load, PV, and wind disturbances are applied across the four interconnected areas to emulate realistic operating conditions for evaluating the proposed GITSMC. Each disturbance includes added zero-mean constant variance Gaussian noise.}
	\label{C1Fig1}
\end{figure}
\begin{figure}[h]
	\hspace{-0.5cm}
	\vspace{-0.1cm}
	\centering
	\includegraphics[width=0.72\textwidth]{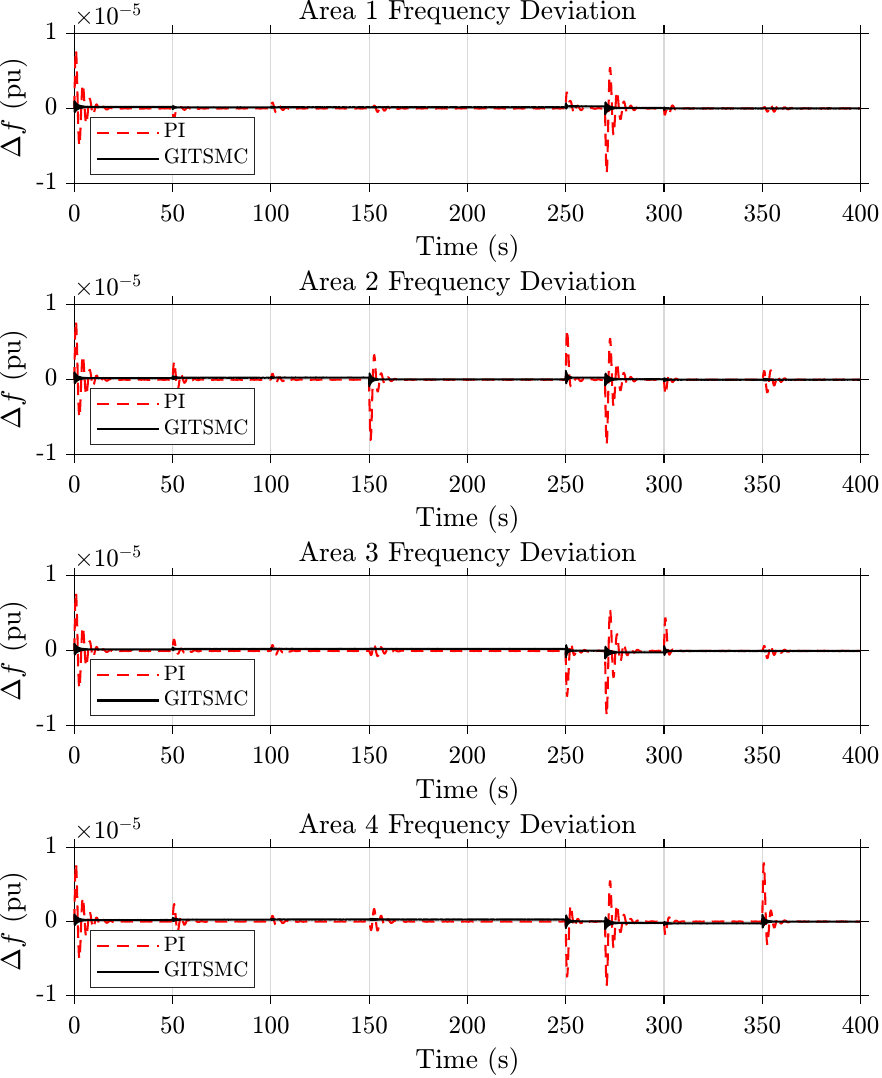}
	\caption{Frequency deviations under variable PV, WT power, and step load fluctuations. The proposed GITSMC shows faster settling and smaller overshoot/undershoot compared to the PI controller.}
	\label{Fig4}
\end{figure}
 \begin{figure}[h]
 		\hspace{-0.5cm}
 	\vspace{-0.1cm}
 	\centering
 	\includegraphics[width=0.72\textwidth]{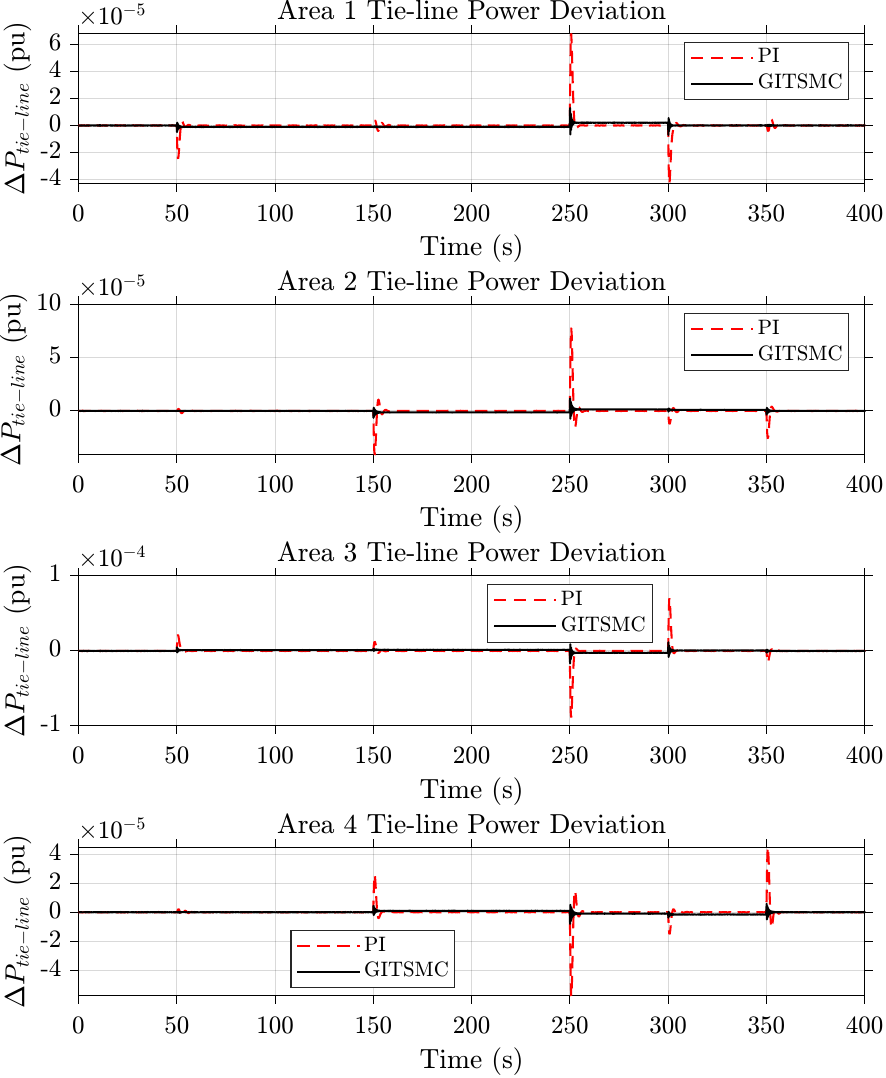}
 	\caption{Tie-line power deviations under load, PV, and WT disturbances. The proposed GITSMC stabilizes the tie-line power quicker and decreases the oscillations compared to the PI controller.}
 	\label{Fig5}
 \end{figure}
The performance of the controllers is evaluated under step load disturbances applied to all four areas. Specifically, a 0.5 pu step load is introduced in area~1 at $t=50$~s and removed at $t=250$~s. In area~2, a 1 pu step load is applied at $t=150$~s and cleared at $t=250$~s, while in areas~3 and 4, a 1 pu step load is applied at $t=250$~s and terminated at $t=300$~s and $t=350$~s, respectively, as illustrated in Fig.~\ref{C1Fig1}. In addition, the PV and WT generation variations, shown in Fig.~\ref{C1Fig1}, are considered in all four areas. The PV power remains at 0 pu from $t=0$--$50$~s and $t=300$--$400$~s, and increases to 0.25 pu during $t=50$--$300$~s. The wind power is 0.8 pu for $t=0$--$100$~s, rises to 0.9 pu during $t=100$--$270$~s, and drops to 0 pu for $t=270$--$400$~s.


\textit{\textbf{Remark 2.} Since variations in WT output power, PV generation, and load demand naturally occur in practical power systems, these disturbances are incorporated to rigorously assess the effectiveness of the proposed controller.}

The frequency deviation and tie-line power
for all areas are shown in Figure~\ref{Fig4} and Figure~\ref{Fig5}, respectively. Figure~\ref{Fig4} illustrates the effect of step load disturbances on the system frequency. When the load increases, the frequency drops noticeably. The proposed GITSMC controller performs better than the classical PI controller in managing these disturbances. It achieves smaller overshoot and undershoot, along with a faster settling time, showing its improved ability to reduce the effects of sudden load changes, variable PV, and WT power. Similar behavior is seen in the tie-line power deviation in Figure~\ref{Fig5}. With the proposed control method, the tie-line power quickly stabilizes, and the oscillations are effectively reduced during load, PV, and WT disturbances. 

The integral performance indices, given in Appendix C, reveal that the proposed GITSMC controller outperforms the conventional PI controller. Although the GITSMC controller achieves ITAE and IAE values of $3.144011 \times 10^{-1}$ and $1.499759 \times 10^{-3}$, which are slightly larger than those of the PI controller ($3.116500 \times 10^{-1}$ and $1.379475 \times 10^{-3}$), the differences remain within acceptable limits, indicating comparable transient performance. While the ITAE and IAE values for the GITSMC controller are slightly higher than those of the PI controller, it shows marked improvement in squared-error metrics, with integral time-weighted squared error (ITSE) and integral squared error (ISE) values of $4.430051 \times 10^{-7}$ and $1.914747 \times 10^{-9}$, respectively, compared to $8.712740 \times 10^{-6}$ and $3.441012 \times 10^{-8}$ for the PI controller. These outcomes unequivocally present that the proposed GITSMC provides pronounced reductions in squared-error energy, indicative of more effective damping, suppressed oscillations, and reinforced disturbance rejection, despite a modest rise in the integral of absolute errors.

\section{Conclusion \& Future Research}
We propose a global integral terminal sliding mode controller (GITSMC) to enhance load frequency control in multi-area interconnected power systems with PV and wind energy integration. The proposed controller employs a global integral terminal sliding surface to achieve finite-time convergence and robust performance under dynamic operating conditions. An exponential reaching law is introduced to suppress chattering, and Lyapunov-based analysis confirms the overall system stability. Simulation results under step load and renewable power variations demonstrate that the proposed controller outperforms the classical PI controller, achieving quicker settling time with decreased overshoot and undershoot in frequency and tie-line power responses. Furthermore, the GITSMC achieves approximately 94.9\% enhancement in ITSE and 94.4\% enhancement in ISE, validating its superior accuracy, robustness, and adaptability in modern power systems.

Future work may focus on developing robust adaptive and virtual inertia controllers to enhance stability under high renewable penetration. Additionally, incorporating machine learning–based and quantum computing–based controllers could further enhance adaptability and optimization in complex power systems. We will also validate the implementation of our controller with a real-world equivalent hardware-in-the-loop simulation in laboratory settings, with a goal of field testing at a later stage.

 

\newpage

\bibliographystyle{unsrt}  
\bibliography{l4dc2026-sample}

@article{islam2025enhanced,
  title={Enhanced Transient Stability in Hybrid {DC/AC} Microgrids: Robust Composite Control Strategy With Virtual Capacitors Integration Using {ANFIS}-Optimized Control Gain Parameters},
  author={Islam, Md Saiful and Bushra, Israt Jahan and Roy, Tushar Kanti and Oo, Amanullah Maung Than},
  journal={IET Renewable Power Generation},
  volume={19},
  number={1},
  pages={e70066},
  year={2025},
  publisher={Wiley Online Library}
}

@article{islam2025marine,
  title={Marine Predators Algorithm-Based Robust Composite Controller for Enhanced Power Sharing and Real-Time Voltage Stability in {DC--AC} Microgrids},
  author={Islam, Md Saiful and Roy, Tushar Kanti and Bushra, Israt Jahan},
  journal={Algorithms},
  volume={18},
  number={8},
  pages={531},
  year={2025},
  publisher={MDPI}
}

@article{islam2025stability,
  title={Stability Enhancement of {DC} Microgrids Under {CPL}s Using Secretary Bird Optimization Algorithm-Tuned Backstepping-{GITSM} Control: Design, Simulation, and Experimental Approach},
  author={Islam, Md Saiful and Bushra, Israt Jahan and Roy, Tushar Kanti and Chowdhury, Jim Mortaej},
  journal={IET Power Electronics},
  volume={18},
  number={1},
  pages={e70118},
  year={2025},
  publisher={Wiley Online Library}
}

@article{islam2024enhanced,
  title={Enhanced composite controller for {PV/PMSG/PEMFC} and {BESS-based DC} microgrids voltage regulation: Integrating integral terminal sliding mode controller and recursive backstepping controller},
  author={Islam, Md Saiful and Bushra, Israt Jahan and Roy, Tushar Kanti and Ghosh, Subarto Kumar and Oo, Amanullah Maung Than},
  journal={IET Generation, Transmission \& Distribution},
  volume={18},
  number={22},
  pages={3608--3632},
  year={2024},
  publisher={Wiley Online Library}
}

@article{feng2025online,
  title={Online Event-Triggered Switching for Frequency Control in Power Grids With Variable Inertia},
  author={Feng, Jie and Cui, Wenqi and Cort{\'e}s, Jorge and Shi, Yuanyuan},
  journal={IEEE Transactions on Power Systems},
  year={2025},
  publisher={IEEE}
}

@article{ghatuari2025coordinated,
  title={A Coordinated Control Strategy of Electric Vehicles for Frequency Control in Modern Power Grids},
  author={Ghatuari, Itishree and Kumar, N Senthil},
  journal={IEEE Access},
  year={2025},
  publisher={IEEE}
}

@article{shangguan2025dissipativity,
  title={Dissipativity-Based Integral-Sliding-Mode Load Frequency Control Considering Disturbances and Denial-of-Service Attacks},
  author={Shangguan, Xing-Chen and Yang, Yuan-Hang and He, Yong and Wei, Chen-Guang and Zhang, Chuan-Ke and Jiang, Lin},
  journal={IEEE Transactions on Power Systems},
  year={2025},
  publisher={IEEE}
}

@article{biswas2025dynamic,
  title={Dynamic Surface Sliding Mode Control-Based {LFC} Design for {RES}-Dominated Power Systems With a Provision of Grid-Scale Virtual Energy Storage},
  author={Biswas, Dip Kumar and Debbarma, Sanjoy and Singh, Piyush Pratap},
  journal={IEEE Transactions on Power Systems},
  year={2025},
  publisher={IEEE}
}

@article{yu2025predictor,
  title={Predictor-Based Fractional-Order Sliding Mode {LFC} for Interconnected Power Systems With Input Delay},
  author={Yu, Xue and Wang, Gang and Zhong, Yuan and Zhang, Huaguang and Liu, Jinhai},
  journal={IEEE Transactions on Cybernetics},
  year={2025},
  publisher={IEEE}
}

@article{zhang2025coordinated,
  title={A Coordinated Adaptive {SMC} Method for Frequency Regulation Control in Power Systems with Multiple Wind Farms},
  author={Zhang, Nan and Zhang, Zheren and Xu, Zheng},
  journal={IEEE Transactions on Sustainable Energy},
  year={2025},
  publisher={IEEE}
}

@article{qian2025load,
  title={Load Frequency Control of Renewable Energy Power Systems Based on Adaptive Global Fast Terminal Sliding Mode Control},
  author={Qian, Jiaming and Lv, Xinxin},
  journal={Applied Sciences},
  volume={15},
  number={13},
  pages={7030},
  year={2025},
  publisher={MDPI}
}

@article{ngamroo2016design,
  title={Design of optimal {SMES} controller considering {SOC} and robustness for microgrid stabilization},
  author={Ngamroo, Issarachai and Vachirasricirikul, Sitthidet},
  journal={IEEE Transactions on Applied superconductivity},
  volume={26},
  number={7},
  pages={1--5},
  year={2016},
  publisher={IEEE}
}

@article{alhassan2025influence,
  title={The Influence of Higher-Order Disturbance Estimation on Wind Power Generation of {WECS} Using {SMC} With Sensorless Wind Speed Estimation},
  author={Alhassan, Ahmad Bala and Do, Ton Duc},
  journal={IEEE Access},
  year={2025},
  publisher={IEEE}
}

@article{sun2017robust,
  title={Robust {H$\infty$} load frequency control of multi-area power system with time delay: a sliding mode control approach},
  author={Sun, Yonghui and Wang, Yingxuan and Wei, Zhinong and Sun, Guoqiang and Wu, Xiaopeng},
  journal={IEEE/CAA Journal of Automatica Sinica},
  volume={5},
  number={2},
  pages={610--617},
  year={2017},
  publisher={IEEE}
}

@article{guo2025observer,
  title={Observer-based compensation control for nonlinear interconnected power systems with load disturbances and actuator faults},
  author={Guo, Bin and Dian, Songyi and Niu, Ben and Zhu, Yuqi and Zhao, Tao},
  journal={IEEE Transactions on Automation Science and Engineering},
  year={2025},
  publisher={IEEE}
}

@article{guha2023improved,
  title={Improved fractional-order sliding mode controller for frequency regulation of a hybrid power system with nonlinear disturbance observer},
  author={Guha, Dipayan and Roy, Provas Kumar and Banerjee, Subrata},
  journal={IEEE Transactions on Industry Applications},
  volume={59},
  number={4},
  pages={4964--4979},
  year={2023},
  publisher={IEEE}
}

@article{yang2021disturbance,
  title={Disturbance observer based fractional-order integral sliding mode frequency control strategy for interconnected power system},
  author={Yang, Fan and Shao, Xinyi and Muyeen, SM and Li, Dongdong and Lin, Shunfu and Fang, Chen},
  journal={IEEE Transactions on Power Systems},
  volume={36},
  number={6},
  pages={5922--5932},
  year={2021},
  publisher={IEEE}
}

@inproceedings{IEEE39bus,
  title={Real-time simulation of {IEEE} 10-generator 39-Bus system with power system stabilizers on miniature full spectrum simulator},
  author={Jain, SK and Narayanan, G and others},
  booktitle={2019 IEEE International Conference on Sustainable Energy Technologies and Systems (ICSETS)},
  pages={161--166},
  year={2019},
  organization={IEEE}
}

@article{jin2023delay,
  title={Delay-dependent stability of load frequency control with adjustable computation accuracy and complexity},
  author={Jin, Li and He, Yong and Zhang, Chuan-Ke and Jiang, Lin and Yao, Wei and Wu, Min},
  journal={Control Engineering Practice},
  volume={135},
  pages={105518},
  year={2023},
  publisher={Elsevier}
}

@book{machowski1997power,
  title={Power system dynamics and stability},
  author={Machowski, Jan and Bialek, Janusz W and Bialek, Janusz and Bumby, James Richard},
  year={1997},
  publisher={John Wiley \& Sons}
}

@misc{John_2017, title={Operation and control of multi-area multi-microgrid systems}, url={https://hdl.handle.net/10356/72682}, DOI={10.32657/10356/72682}, abstractNote={There has been a widespread deployment of microgrids around the world in recent years. Microgrids form a local area power distribution system with distributed generations, energy storage systems and controllable loads. The next stage of innovation in the field of microgrid systems is the interconnection of several AC and DC microgrid systems spread over large geographical distances to form multi-area multi-microgrid (MMG) systems, which will satisfy the ever increasing global energy demands. There are several benefits associated with these MMG systems such as improved reliability and security of power supply, mutual power sharing and reduced investment in new generating capacity. For the effective operation and control of these MMG systems during different modes of operation, effective methods of power, voltage and frequency controls are essential. 

First, different system architectures are proposed for AC/AC multi-area MMG system, consisting of interconnected AC microgrids and for AC/DC multi-area MMG system, consisting of both interconnected AC and DC microgrids. Control systems consisting of centralized and local controllers are proposed for the effective control of load bus voltages and frequency, inverter and converter power outputs, and power exchange between the interconnected microgrids in each of these MMG systems. The local control of the converters and inverters in these MMG systems is realized using a state-space model based control algorithm, namely model predictive control (MPC). The proposed MPC algorithm, unlike the existing MPC algorithms, is independent of grid, line and load impedances in the MMG systems. In comparison with conventional proportional-integral (PI) control methods, the proposed MPC algorithm gives smaller tracking error, shorter settling time, and better steady-state and transient responses in different operating modes of the interconnected microgrids such as grid-connected and islanded modes. In addition to the system architectures and control systems, different load shedding schemes are also proposed for the MMG systems. An underfrequency load shedding scheme is proposed for the AC/AC multi-area MMG system for effective voltage and frequency regulation during AC microgrid islanding. Also, an undervoltage load shedding scheme is proposed for the AC/DC multi-area MMG system for effective voltage and power regulation during DC microgrid islanding.     

Then, various simulation studies are conducted to test the operation and control of these MMG systems under different operating conditions such as microgrid islanding, power exchanges, load changes, load shedding and line outages. The simulation studies show that the developed control systems in these MMG systems can achieve good control performance and effective voltage, frequency and power regulation under different operating conditions.  Thus, the effectively controlled multi-area MMG systems are capable of fulfilling basic objectives such as improved reliability and security of power supply, enhanced voltage and frequency stability, and effective dynamic islanding.

Finally, to solve the various power quality issues such as current distortion, voltage distortion, voltage sag, voltage unbalance and low power factor in the proposed AC/AC multi-area MMG system, a new power quality improvement method is proposed. A new method of non-local harmonic current and reactive power compensation using a series-shunt network device (SSND) is proposed for the AC/AC multi-area MMG system. Even though local harmonic current and reactive power compensation methods are available, non-local compensation methods, based on their several advantages, are alternatives to be necessarily considered in the future for large MMG systems, which consist of widely dispersed loads. SSND consisting of series and shunt inverters is installed in the line interconnecting two microgrids in the AC/AC multi-area MMG system. A state-space model based MPC algorithm is used for the proposed power quality improvement scheme to regulate various parameters such as output voltage, frequency, current and power of multiple inverters and converters in the MMG system integrated with the SSND. The power flow and power quality control functions of the SSND are analyzed theoretically to understand the different capabilities of SSND in harmonic current and reactive power compensation and in voltage disturbance isolation in the MMG system. 

Several simulation studies are conducted to demonstrate the effective operation of SSND using MPC in the proposed AC/AC multi-area MMG system. From these simulation studies, it is verified that SSND can effectively achieve local and non-local harmonic current and reactive power compensation, and can also isolate one microgrid from voltage disturbances such as voltage distortion, voltage sag and voltage unbalance occurring in the adjacent microgrid. In addition, SSND can provide emergency real power support during islanding of a microgrid in the MMG system.}, author={John, Thomas}, year={2017}}

@inproceedings{roy2024multi,
  title={Multi-Area Load Frequency Control Using an Adaptive Reaching Law-Based Integral Terminal Sliding Mode Scheme},
  author={Roy, Tushar Kanti and Mahmud, Md Apel and Oo, Amanullah Maung Than},
  booktitle={2024 IEEE International Conference on Power Electronics, Drives and Energy Systems (PEDES)},
  pages={1--6},
  year={2024},
  organization={IEEE}
}

@article{roy2024robust,
  title={Robust {LFC} design using adaptive neuro-fuzzy inference-aided optimal fractional-order {PIDA} control for perturbed power systems with solar and wind power sources},
  author={Roy, Tushar Kanti and Yu, Samson S and Mahmud, Md Apel and Trinh, Hieu},
  journal={IET Generation, Transmission \& Distribution},
  volume={18},
  number={12},
  pages={2193--2212},
  year={2024},
  publisher={Wiley Online Library}
}

@article{sahu2015design,
  title={Design and analysis of hybrid firefly algorithm-pattern search based fuzzy {PID} controller for {LFC} of multi area power systems},
  author={Sahu, Rabindra Kumar and Panda, Sidhartha and Pradhan, Pratap Chandra},
  journal={International Journal of Electrical Power \& Energy Systems},
  volume={69},
  pages={200--212},
  year={2015},
  publisher={Elsevier}
}

@inproceedings{sharifi2008load,
  title={Load frequency control in interconnected power system using multi-objective {PID} controller},
  author={Sharifi, A and Sabahi, K and Shoorehdeli, M Aliyari and Nekoui, MA and Teshnehlab, M},
  booktitle={2008 IEEE conference on soft computing in industrial applications},
  pages={217--221},
  year={2008},
  organization={IEEE}
}

@article{yan2020multi,
  title={A multi-agent deep reinforcement learning method for cooperative load frequency control of a multi-area power system},
  author={Yan, Ziming and Xu, Yan},
  journal={IEEE Transactions on Power Systems},
  volume={35},
  number={6},
  pages={4599--4608},
  year={2020},
  publisher={IEEE}
}
\appendix
\section{Finite Time Convergence Analysis}
\label{app:A}
\renewcommand{\theequation}{A.\arabic{equation}}
\setcounter{equation}{0}

The finite convergence time $(t_f)$ of the GITSMC controller for a nonzero initial state $x_i(0)$ is determined as follows. The sliding mode occurs when $\Theta_i(t) = 0$. From Equation~\eqref{eq:CD1}, the system dynamics can be expressed as:
\begin{eqnarray}
\begin{aligned}\label{eq:CD15}
\dot{x}_{i}(t) = -\lambda{i}x_{i}^{\alpha_{i}}(t)
\end{aligned}
\end{eqnarray}
Rearranging:
\begin{eqnarray}
\begin{aligned}\label{eq:CD16}
dt = -\frac{dx_{i}}{\lambda_{i}x_{i}^{\alpha_{i}}}
\end{aligned}
\end{eqnarray}
Integrating:
\begin{eqnarray}
\begin{aligned}\label{eq:CD17}
\int_{0}^{t_f} dt = -\int_{x_i(0)}^{x_i(t_f)} \frac{dx_{i}}{\lambda_{i}x_{i}^{\alpha_{i}}}
\end{aligned}
\end{eqnarray}
Solving Equation~\eqref{eq:CD17} provides the finite-time convergence expression as:
\begin{eqnarray}
\begin{aligned}\label{eq:CD18}
t_f = \frac{|x_i(t_f)|^{1-\alpha_i} - |x_i(0)|^{1-\alpha_i}}{\lambda_i(\alpha_i - 1)}
\end{aligned}
\end{eqnarray}
Hence, the system trajectory reaches the origin within a finite time, confirming the finite-time stability property.

\section{State-Space Matrices of the Four-Area Power System}
\label{app:B}
Based on the parameters provided in Table~\ref{tab2}, the state-space representations for the four areas are obtained as follows.

\textbf{Area 1:}
\[
A_{01} =
\begin{pmatrix}
0 & 8.33 & 0 & 0 & 0 & 0 & 0\\
-5 & -5 & 5 & 0 & 0 & 5 & 5\\
0 & 0 & -2.67 & 0 & 2.67 & 0 & 0\\
5 & 11.15 & 0 & 0 & 0 & 0 & 0\\
0 & -264.07 & 0 & -12.43 & -12.43 & 0 & 0\\
0 & 0 & 0 & 0 & 0 & -25 & 0\\
0 & 0 & 0 & 0 & 0 & 0 & -23.81
\end{pmatrix},\quad 
B_{01} = [0~~0~~0~~0~~12.43~~0~~0]^T,
\]
\[
\vartheta_1 = [0~~0~~0~~0~~1/12.43~~0~~0].
\]

\textbf{Area 2:}
\[
A_{02} =
\begin{pmatrix}
0 & 5.70 & 0 & 0 & 0 & 0 & 0\\
-3.33 & -3.33 & 3.33 & 0 & 0 & 3.33 & 3.33\\
0 & 0 & -2.65 & 0 & 2.65 & 0 & 0\\
5 & 99.69 & 0 & 0 & 0 & 0 & 0\\
0 & -279.36 & 0 & -13.15 & -13.15 & 0 & 0\\
0 & 0 & 0 & 0 & 0 & -25 & 0\\
0 & 0 & 0 & 0 & 0 & 0 & -23.81
\end{pmatrix},\quad 
B_{02} = [0~~0~~0~~0~~13.15~~0~~0]^T,
\]
\[
\vartheta_2 = [0~~0~~0~~0~~1/13.15~~0~~0].
\]

\textbf{Area 3:}
\[
A_{03} =
\begin{pmatrix}
0 & 12.68 & 0 & 0 & 0 & 0 & 0\\
-2.92 & -2.92 & 2.92 & 0 & 0 & 2.92 & 2.92\\
0 & 0 & -2.58 & 0 & 2.58 & 0 & 0\\
5 & 107.88 & 0 & 0 & 0 & 0 & 0\\
0 & -423.37 & 0 & -13.33 & -13.33 & 0 & 0\\
0 & 0 & 0 & 0 & 0 & -25 & 0\\
0 & 0 & 0 & 0 & 0 & 0 & -23.81
\end{pmatrix},\quad 
B_{03} = [0~~0~~0~~0~~20.57~~0~~0]^T,
\]
\[
\vartheta_3 = [0~~0~~0~~0~~1/20.57~~0~~0].
\]

\textbf{Area 4:}
\[
A_{04} =
\begin{pmatrix}
0 & 6.33 & 0 & 0 & 0 & 0 & 0\\
-3.22 & -3.22 & 3.22 & 0 & 0 & 3.22 & 3.22\\
0 & 0 & -2.45 & 0 & 2.45 & 0 & 0\\
5 & 107.66 & 0 & 0 & 0 & 0 & 0\\
0 & -257.31 & 0 & -12.53 & -12.53 & 0 & 0\\
0 & 0 & 0 & 0 & 0 & -25 & 0\\
0 & 0 & 0 & 0 & 0 & 0 & -23.81
\end{pmatrix},\quad 
B_{04} = [0~~0~~0~~0~~12.53~~0~~0]^T,
\]
\[
\vartheta_4 = [0~~0~~0~~0~~1/12.53~~0~~0].
\]

\appendix
\section*{Appendix C. Integral Indices}
\label{app:CC}
\renewcommand{\theequation}{C.\arabic{equation}}
\setcounter{equation}{0}

The following integral indices are utilized to evaluate the dynamic performance of the proposed controller in terms of frequency and tie-line power deviations across the interconnected areas~\citep{roy2024robust, roy2024multi}.

\begin{equation}
{\mathrm{ITAE}} = \int_{0}^{t} 
\left(
\sum_{i=1,\, i \ne j}^{n} 
\left( |\Delta f_i| + |\Delta P_{\mathrm{tie\text{-}line}\, ij}| \right)
\right)
t \, d t
\end{equation}

\begin{equation}
{\mathrm{ITSE}} = \int_{0}^{t} 
\left(
|\Delta f_i|^2 + |\Delta P_{\mathrm{tie\text{-}line}\, ij}|^2
\right)
t \, d t
\end{equation}

\begin{equation}
{\mathrm{ISE}} = \int_{0}^{t} 
\left(
|\Delta f_i|^2 + |\Delta P_{\mathrm{tie\text{-}line}\, ij}|^2
\right)
d t
\end{equation}

\begin{equation}
{\mathrm{IAE}} = \int_{0}^{t} 
\left(
\sum_{i=1,\, i \ne j}^{n} 
\left( |\Delta f_i| + |\Delta P_{\mathrm{tie\text{-}line}\, ij}| \right)
\right)
d t
\end{equation}

\end{document}